# An optimal scheduling architecture for accelerating batch algorithms on Neural Network processor architectures

Phani Kumar Nyshadham, Sinha Mohit, Mishra Biswajit, H S Vijay

Intel Corporation, Santa Clara, USA

## ABSTRACT

In neural network topologies, there are algorithms running on batches of data tensors. The batches of data are typically scheduled onto the computing cores which execute in parallel. For the algorithms running on batches of data, an optimal batch scheduling architecture is very much needed by suitably utilizing hardware resources - thereby resulting in significant reduction training and inference time. In this paper, we propose to accelerate the batch algorithms for neural networks through a scheduling architecture enabling optimal compute power utilization. The proposed optimal scheduling architecture can be built into HW or can be implemented in SW alone which can be leveraged for accelerating batch algorithms. The results demonstrate that the proposed architecture speeds up the batch algorithms compared to the previous solutions. The proposed idea is applicable to any HPC architecture meant for neural networks.

## 1. INTRODUCTION

The algorithms running on batches of data are often found in Attention based Neural Network like BERT [1], Transformer [2]. The data tensors in each batch are loaded into device memory of a HPC neural network processor. Generally, the tensors in all the batches is organized into N-dimensional tensor as a whole. The N-dimensional tensor will of course be laid out as 2D tensor in physical memory with demarcations for batches as per the engineering convenience.

For a set of input features, there are a sequence of algorithms running in a neural network from the input layer through the hidden layers to the output layer. For each iteration, let it be the forward pass or the backward pass, there are algorithms running on batches of data tensors. Typically, the data to be processed in each batch is small but number of batches can be in the order of hundreds. In a neural network there are several such operations that make it computationally very expensive to train and infer.

There have been conventional techniques to schedule the batches onto the computational cores so that they execute in parallel. Some of these scheduling architectures are not optimal because they suffer from the problem of compute resource under-utilization. In this paper, we address the problem of compute resource under-utilization to accelerate the batch algorithms for HPC neural network architectures.

## 2. PROBLEM STATEMENT

For example, in BERT topology, algorithms like SOFTMAX are supposed to run on batches of data tensors (for example, tensor dimensions in every batch can be 512 x 512) and the number of batches can be 384 (as an example). If all the batches in the example are considered as 3D tensor (as a whole), the dimensions of the whole tensor in 2D layout are 512 * 384 x 512. In the above example, running SOFTMAX on every slice of 512 x 512 inside the whole tensor (512 * 384 x 512) turns out to be underperforming for the following reasons:

1. The slices can be small enough to be split across all the compute clusters

2. HPC neural network architectures has a *concept of nearness* of the Compute clusters/cores to the banks of device memory which has a direct impact on load-store latency which means that a *compute cluster/core has the least possible latency only when it accesses data (load/store) from a nearest memory bank compared to other memory banks*.

Given the above reasons, the HW utilization would suffer when:
- ✓ Creation of the whole tensor in the device memory is not concentrated but rather distributed across banks of device memory
- ✓ Data tensor distribution cannot be uniform always across the banks of device memory

## 2.1 Cluster-memory bank nearness posing challenges to the scheduling architecture:

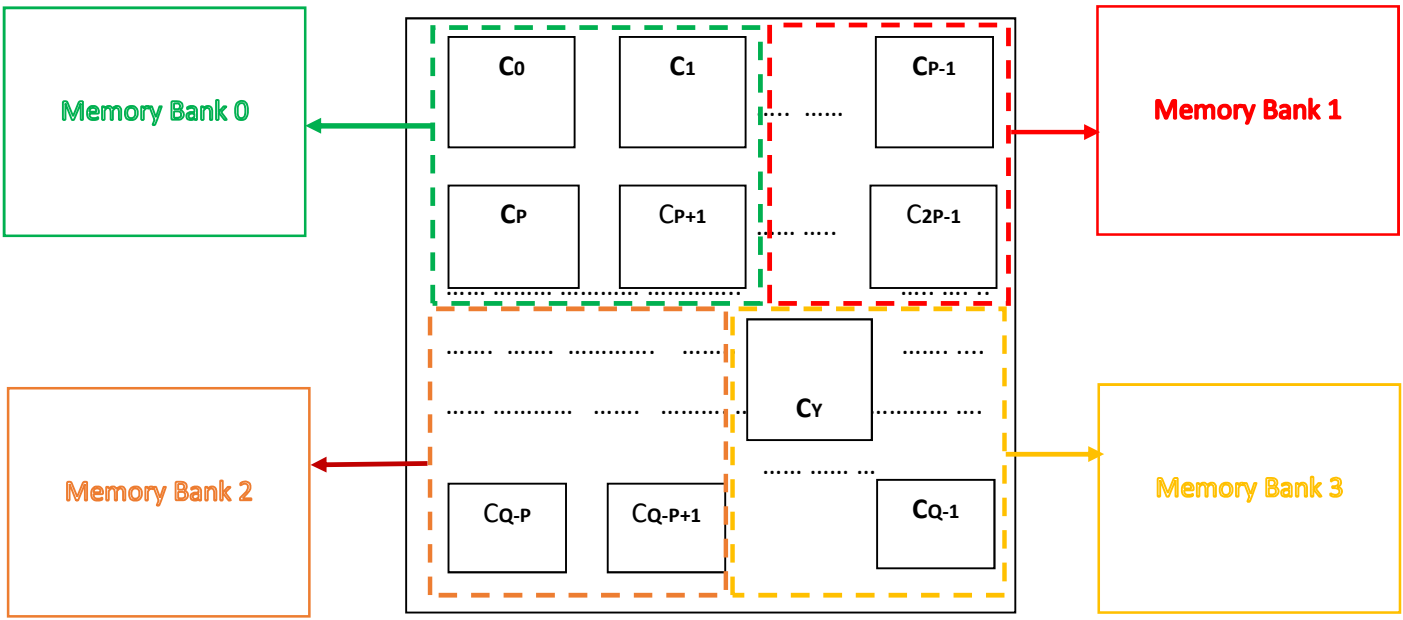

The above data-layout diagram explains the concept of concept of nearness of the Compute clusters/cores to the banks of device memory. The above diagram depicts a common HPC processor architecture for neural network platform where there are Q compute clusters and there are 4 banks of memory. The Clusters inside a dotted color box are all grouped and associated with *near-by* memory bank with the same color for least possible latency for load/store operations. If the clusters access the memory banks with a different color, the latency is more for load/store operations.

For example, $C_Y$ has least latency when it accesses memory bank-3 while it would incur huge latency while accessing memory bank-0/1/2. Not all farther banks have same latency but more latency than the nearest associated bank. The latency between a cluster and a farther memory-bank depends on the number of hops it takes to reach that bank.

With improper scheduling of batches onto the compute clusters/cores which affects the cluster-memory bank nearness aspects, the batch algorithms suffer from severe under-utilization of compute power on HPC neural network architectures. In this paper, we address the problem of compute resource under-utilization resulting from improper work-load scheduling to accelerate the batch algorithms for neural networks on HPC architectures.

## 3. PREVIOUS SOLUTIONS

A common way of running algorithms on batches of data is to repeatedly run the algorithm on each batch and trying to utilize all the computing clusters in parallel. Many times, when each batch of data tensor is scheduled onto all the compute clusters/cores, the workload in every batch can be small resulting in compute power under-utilization.

To solve the compute utilization problem in the earlier case, instead of trying to schedule each batch onto all the available compute clusters/cores, every batch is scheduled onto one compute cluster/core not all. Commonly, round-robin scheduling (RR) is followed for scheduling the batches onto the clusters/cores. This has certainly alleviated the resource under-utilization but not to the best possible extent. RR performs optimally only when the whole tensor is on one bank of device memory.
This round-robin (RR) scheduling of batches onto the compute clusters/cores proved to be sub-optimal and could only improve the utilization by marginal amount. At times, severe latency would result because of memory-bank and compute cluster/core nearness aspects.

## 4. PROPOSED SOLUTION

In this paper, we address the problem of compute resource under-utilization to accelerate the batch algorithms for neural networks. We propose to accelerate the batch algorithms for neural networks through an architecture (HW/SW) to achieve optimal compute power utilization. We use **batch (or) patch** throughout the document which refers to a slice of whole tensor which refers to the basic unit of workload for batch algorithms. In the rest paper, we use the terms **clusters and cores** interchangeably which mean the same and refers to a basic computing unit inside a multi-processor. This optimal compute resource utilization architecture consists of 2 steps as follows:

### 4.1 Compute Cluster beamforming algorithm

By *Compute Cluster Beamforming*, we mean every batch of data tensor is to be steered onto a compute cluster such that the scheduling is optimal. Instead of RR scheduling onto all the clusters, we propose Cluster Beamforming algorithm as explained below.

For every batch of a data tensor, we look for the bank of device memory where it is created. A table is to be constructed which enumerates the information of batch-to-bank mapping as depicted below. In the below table, batch $P_0$ is actually created in $H_z$-bank of device memory. Similarly, $P_1$ is created on $H_x$-bank. When we observe $P_{N-2}$, a portion of it is created on $H_y$-bank and the other portion on $H_z$-bank. That situation can arise especially when the tensors are created on the boundaries of device memory banks. That's why OVERLAP field is set in the table for $P_{N-2}$. Thus, the whole table is constructed for $N$-batches of data tensors.

From the table, we can see that $\{N_x, N_y, N_z, N_w\}$ batches are created on $\{H_x, H_y, H_z, H_w\}$ banks respectively. Because of OVERLAP fields being set for some batches,
$$N_x + N_y + N_z + N_w > N$$

|     | Hx | Hy | Hz | Hw | Overlap |
| --- | --- | --- | --- | --- | --- |
| $P_0$ | 0 | 0 | 1 | 0 | 0 |
| $P_1$ | 1 | 0 | 0 | 0 | 0 |
| …. | …. | …. | …. | …. | …. |
| $P_{N-2}$ | 0 | 1 | 1 | 0 | 1 |
| $P_{N-1}$ | 0 | 0 | 0 | 1 | 0 |
|     | $N_x$ | $N_y$ | $N_z$ | $N_w$ |     |

For the batches with OVERLAP field, the algorithm checks for the proportion of the batch created across banks and assigns the batch to the bank with maximum proportion of created tensor. In case of equal proportions, the banks are chosen with equal probability. After this step, a new set of batches {*Lx, Ly, Lz, Lw*} are assigned to {*Hx, Hy, Hz, Hw*} banks respectively such that

$$Lx + Ly + Lz + Lw = N$$

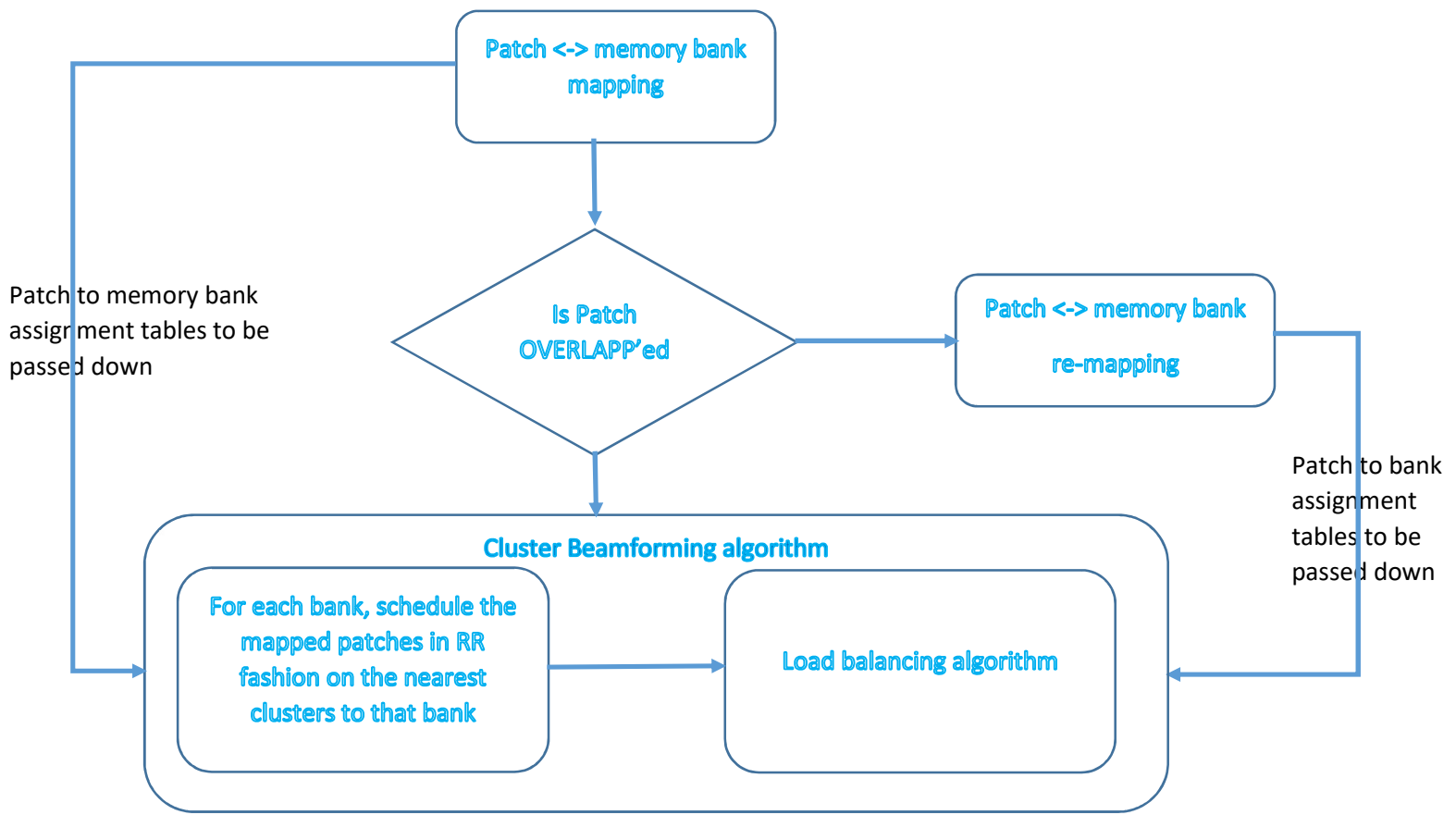

*Fig-1: Flowchart of Cluster Beamforming concept*

Of course, there can be other variations to the assignment of OVERLAPP'ed batches to the banks based on assigned load on banks apriori.

We iterate over the banks such that, for every bank, we schedule the patches created on that bank in a round-robin manner onto the compute clusters nearest to that bank of device memory. This concept takes into consideration the cluster-memory bank nearness aspect of HPC neural network architecture yielding minimum latency for load/store operations. The flowchart for the Cluster Beamforming algorithm is explained in Fig-1.

*Improvement of the proposed Cluster Beamforming concept over conventional RR scheme:*
The proposed Cluster beamforming algorithm optimizes the load-store latency between the compute clusters and the memory-banks considering the associated nearness aspects. HPC neural network architectures are based on the concept of many-to-one mapping of compute clusters to a memory-bank which offers least possible latency for load and store operations. Which means, the compute clusters are not forbidden to access a farther memory bank but with more latency. As the "virtual" distance between a compute cluster and memory bank is more, the latency can be exorbitantly high at times.

The proposed Cluster Beamforming hence proceeds with constructing a mapping table (assignment table) between the list of created patches and the memory banks on the device based on the location of allocated patches on memory banks of device memory. This guarantees that a patch is always scheduled on a compute cluster which is "near" to the memory bank on which the patch is allocated. Once the patches are mapped/assigned to a memory bank, then it can be scheduled onto any compute cluster **"near"** to that memory bank. Thus, the concept is optimal by achieving least possible latency with the proposed scheduling concept.

On the contrary, conventional RR scheduling schedules the patches across the pool of compute clusters. This could result in a patch getting mapped/assigned to a compute cluster which is **"far"** to the memory bank on which it lies. This results in latent load-store operations and hence this approach is sub-optimal in nature.

### 4.2 Load balancing algorithm – Non-uniform tensor distribution

There can be a situation when the whole tensor with all the batches is concentrated on a few banks and is not uniformly distributed across banks. In this case, *Lx, Ly, Lz, Lw* are not equal and some values are very high and some values are very low. This is the problem of **UNBALANCED WORKLOAD** distribution. This can also occur in the cases where each batch of the tensor is distributed across all the memory banks, in which case the patch to memory bank mapping table always looks at the first portion of the patch and assigns all the patches onto one memory bank.

We propose Load balancing algorithm as a second step to complement the Cluster Beamforming algorithm in step-1. In this step, we propose to take off some load (optimal number of patches) from the overloaded banks and *virtually assign* to the under-loaded banks to ensure the load is balanced. After this, the batches are scheduled in RR fashion on the newly assigned banks.

We also propose a weighted load balancing algorithm such that the load is split between the overloaded banks and *virtual* under-loaded banks in the inverse proportion of the latencies. Which means, the clusters farther from the allocated memory bank share less load compared to the clusters closest to the allocated memory bank. It is to be noted that there is no actual re-allocation of patches onto under-loaded banks. That's' why we use "virtual" keyword wherever applicable. The latencies are computed as a function of distance from the overloaded banks to the compute clusters on the under-loaded banks. The inverse ratio of latencies provides the weights to split the load between over-loaded and under-loaded compute clusters. The flowchart for the Load balancing algorithm is explained in Fig-2.

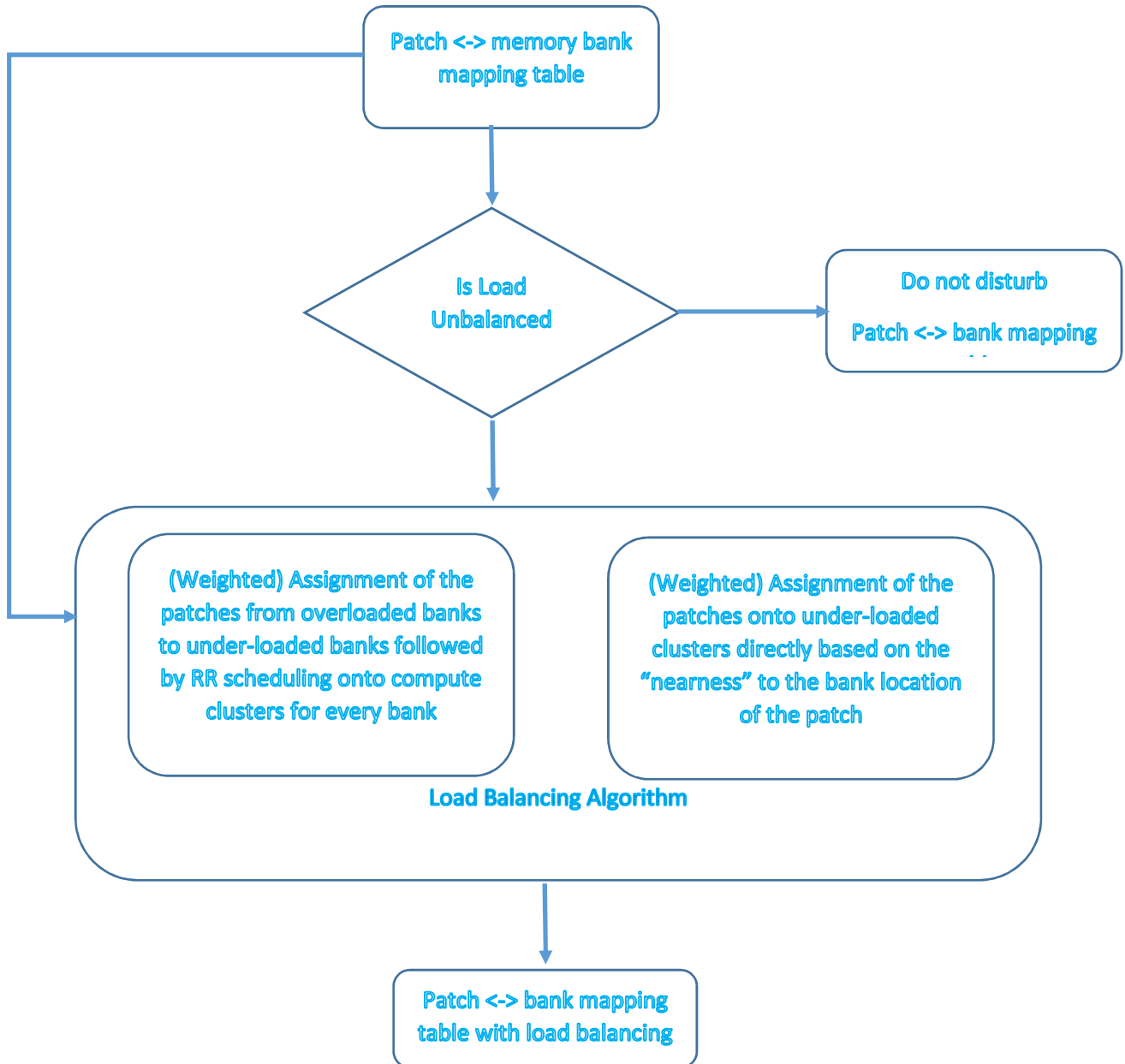

Fig-2: Flowchart of Load Balancing algorithm complementing Cluster Beamforming

There can be other variations to perform workload re-assignment based on compute cluster instead of memory banks. We can take off a part of load from a compute cluster and assign to the under-loaded compute cluster based on the nearness of the under-loaded cluster to the bank. Many other variations are possible based on the following MINIMAX optimization problem to be solved for "optimal" workload assignment $S_i$ on $i^{th} core$, $\{i \in I\}$, $given\ a\ set\ of\ patches\ S$, as explained below:

$$min_{S_i,\{i \in I\}} \left( max_{\{compute\ core_i, p_i \epsilon S_i\}} \left( \sum_{p_i \epsilon S_i} E_t(p_i) \right) \right) ; \cup_{i=0}^{i=I-1} S_i = S; n(S_i) = N_i$$

$\qquad$ where the device has I compute cores
$\qquad E_t(p_i) = $ Execution time of $p_i^{th}$ Patch on compute core $'i' = f(d(p_i, i))$
$\qquad (d(p_i, i)) = $ distance between Patch $p_i$ and compute core $'i'$
$\qquad S_i, \{i \in I\}$ is the set of patches scheduled onto compute core $'i'$
$\qquad S$ is the set of patches scheduled by the batch algorithm
$\qquad N_i$ is the number of patches scheduled onto compute core $'i' = n(S_i)$
$\qquad N$ is the number of patches scheduled onto device with I cores $= n(S)$

## 5. Deployment of the proposed scheduling architecture on HW

The proposed concept of Cluster Beamforming coupled with load balancing can be built into the HW architecture as a controller block for scheduling the workload. This would reduce the SW overhead to schedule the patches onto the compute clusters optimally.

**Example data layout of patches**

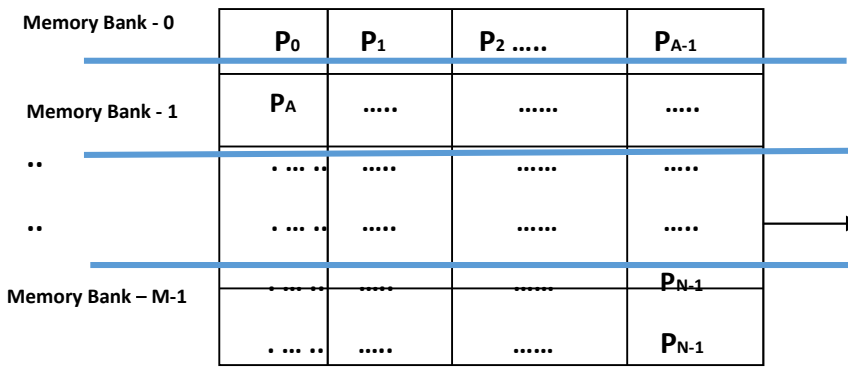

**HW BATCH SCHEDULER**

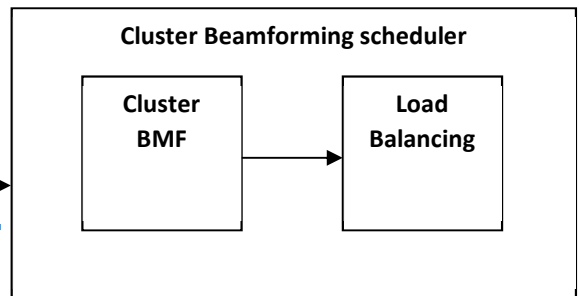

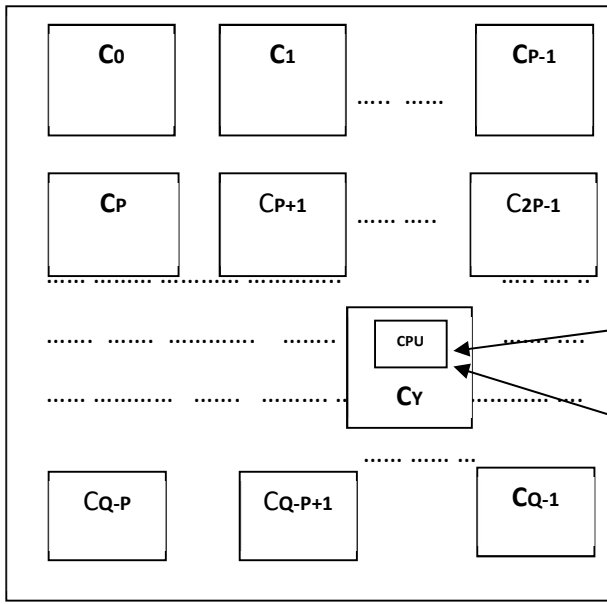

**Multi-Cluster Tensor Processing Unit**

**Patch<->Cluster Mapping Register file**

For any batch algorithm, the whole tensor after creation can be input to the scheduler block inside HW which constructs the table of patch <-> compute cluster mapping upfront. The kernel code would be generated for every patch in templated form and the patch number (patch geometry) would be picked up from the table constructed by the scheduler block for that compute cluster. This way the kernel instructions are generated only once – the same set of instructions would be executed with the patch geometry as variable.

The above block diagram depicts the interplay between the Cluster-CPUs and the HW BATCH SCHEDULER which incorporates the proposed Cluster Beamforming coupled with Load Balancing algorithms. An example data layout of a whole tensor built with *N* patches of smaller tensors is showed in the figure. HW BATCH SCHEDULER is depicted as a module working on patches of tensors and running the Cluster Beamforming and Load Balancing algorithms depicted in the aforementioned flowcharts. Also an example Neural Network Multi-Cluster Tensor Processing Unit for is depicted with *Q* clusters.

The BLUE lines show an example allocation of the tensor across *M* memory banks. The output of HW batch scheduler is **Patch<->Cluster mapping table** which can be stored as a REGISTER file. The geometry information of each of the mapped patches for every cluster is also stored as part of mapping table.

During kernel generation, the CPU of the cluster reads the REGISTER file and provides the geometrical coordinates of the mapped patches onto the cluster in a queue. These coordinates are input to the kernel templated instructions and get executed. In the below figure, for cluster $C_y$, the CPU inside queues up the patches enumerated in REGISTER file onto $C_y$ as $P_x$ and $P_{N-2}$. The geometrical information about these patches ($P_x$ and $P_{N-2}$) is picked up by the CPU of cluster $C_y$ and the kernel executes the templated instructions with the geometrical information for every mapped patch.

## 6. EXPERIMENTAL RESULTS on INTEL NNP-T ACCELERATOR

In this section we provide results of comparison between RR scheduling and proposed concept applied to a couple of batch algorithms for BERT neural network topology.

| Matrix Multiplication | Batch Size | CBMF / RR HW utilization |
|---|---|---|
| (512 x 64) x (64 x 512) | 384 | 3.5x |
| (512 x 512) x (512 x 64) | 384 | 3.35x |

The above table depicts the comparison and improvement of the proposed idea (Cluster beamforming CBMF) with respect to the conventional batch scheduling algorithm (round-robin RR). It is observed that the HW utilization is improved which very much is directly proportional to the algorithm acceleration. The results are tabulated for different dimensions of matrix multiplication.

In the case of SOFTMAX batched algorithm, the below table demonstrates the improvement in HW utilization observed between conventional RR scheduling and proposed CBMF architecture for the below tabulated dimensions. The improvement varies across algorithms based on the nature of the algorithm (memory-bound/compute-bound) and problem sizes. In any case, proposed Cluster Beamforming always performs optimal scheduling.

| SOFTMAX | Batch Size | CBMF / RR HW utilization |
|---|---|---|
| 512 x 512 | 384 | 1.08x |

## 7. CONCLUSION

This paper proposed to accelerate the batch algorithms for neural networks through an architecture enabling optimal compute power utilization through Cluster beamforming technique coupled with a Load balancing algorithm. We propose a design of scheduling architecture which can be built into HW or can be implemented in SW without changes to HW. Either way, the proposed idea is designed for accelerating and executing batch algorithms. The proposed idea is applicable to any HPC architecture meant for neural networks. With the proposed optimal scheduling architecture, the batch algorithms in neural networks, can be accelerated utilizing hardware resources - thereby resulting in significant reduction training and inference time. The proposed idea is applicable to all Tensor Processing & High-Performance Computing architectures.